\documentclass[a4paper,11pt]{article}
\pdfoutput=1

\usepackage{jcappub} 
\usepackage[T1]{fontenc} 
\usepackage{bm}
\usepackage{mathrsfs}
\usepackage{float}

\title{\boldmath A General Model for Dark Energy Crossing the Phantom Divide}

\author[1]{Zhibang Yao,}
\author[1]{Gen Ye}
\author[1]{and Alessandra Silvestri}

\affiliation[1]{Institute Lorentz, Leiden University, PO Box 9506, Leiden 2300 RA, The Netherlands}

\emailAdd{yao@lorentz.leidenuniv.nl}

\abstract{Within the framework of spatially covariant theories, we propose a general model for dark energy (DE) in which the cosmological background and perturbations are independently controlled by different sets of coefficients, and the equation of state of DE is directly determined by two free functions of time from the Lagrangian. These properties allow to realize arbitrary background evolutions while avoiding ghost and gradient instabilities in linear perturbations. They also enable a more direct  analysis of phantom crossing  without having to first solve the background equations of motion. In this model, the sound speed of the scalar mode is scale-dependent and approaches infinity at large scale, so that the field becomes non-dynamical in the infrared (IR) limit. Even though this usually indicates a strong coupling issue, we speculate that this is avoided  because the scalar degree of freedom becomes frozen not only at linear order but also at any higher order in IR limit. Given this characteristic large scales behavior, we dub the model \emph{Freezing Gravity}. On smaller scales, the scalar mode propagates with a finite speed of sound. The theory has a cut-off in energy, signaled by the pole in the speed of sound, when the effective Planck mass exceeds Planck mass.}

\begin{document}
\maketitle
\flushbottom

\section{Introduction \label{sec_Introduction}}

Recent results from the Dark Energy Spectroscopic Instrument (DESI) collaboration~\citep{DESI:2024mwx,DESI:2024aqx,DESI:2024kob} showed a tension between measurements of baryon acoustic oscillations (BAO) and cosmic microwave background (CMB) radiation when analyzed within the Lambda Cold Dark Matter ($\Lambda$CDM) model. This tension further increased to around the 3$\sigma$ level with the latest data releases from the DESI~\citep{Collaboration:2025aa} and the Atacama
Cosmology Telescope (ACT)~\citep{Louis:2025aa} collaborations, and was shown to be alleviated within a $w_{0}w_{a}$ parametrization \citep{DESI:2024aqx,DESI:2024kob,Chevallier:2001aa,Linder:2003aa,Gialamas:2025pwv} for the dark energy (DE) equation of state (EoS). These results favor a time-varying DE and, when combined with the data set of Type Ia supernovae (SNIa) \citep{Collaboration:2024aa}, provide hints for an EoS of DE that crosses $-1$ i.e., the phantom divide \citep{Lodha:2025aa,Ormondroyd:2025aa,Gu:2025aa,Ye:2025ark}. This latest implication in particular, has stimulated a significant investigation to identify robust theoretical embeddings (see e.g.~\cite{Orchard:2024bve,Chudaykin:2024gol, Alestas:2024gxe,Wang:2024dka,Notari:2024rti,Gialamas:2024lyw,Akarsu:2024eoo,Heckman:2024apk,Mukherjee:2024ryz,Giare:2024smz,Berghaus:2024kra,Yin:2024hba,Tada:2024znt,Bhattacharya:2024hep,Scherer:2025esj,Nojiri:2025low,Hogas:2025ahb}), reviving the interest in dynamical dark energy~\citep{Copeland2006}. 

It is well known that phantom crossing cannot be realized in the minimally coupled single scalar field scenario with canonical kinetic term i.e., quintessence, without exciting ghost instabilities \citep{Caldwell:1998aa,Carroll:aa}. In order to achieve that, one may consider multiple scalar fields (e.g. quintom \citep{Feng:2005aa}), nonlinear kinetic terms (e.g. $k$-essence \citep{Vikman:2005aa}), non-minimal coupling (e.g. $f\left(\mathscr{R}\right)$
theory \citep{Gannouji:2006aa,Amendola:2008aa}) or higher order derivative terms (e.g. effective quintessence \citep{Creminelli:2008wc}) as well as other possible modifications (e.g. $f\left(T\right)$ theory \citep{Cai:2016aa}). In a recent work based on nonparametric reconstruction methods~\citep{Ye:2025aa}, it was shown that, within the broad framework of Horndeski gravity, non-minimal coupling is preferred by the DESI results, see also \cite{Wolf:2024stt,Wolf:2025jed}. In the same paper, the authors proposed a concrete model, dubbed \emph{thawing gravity}, as a specific example of non-minimally coupled gravity which fits the DESI results better than $\Lambda$CDM. Common to all these approaches, is the fact that one usually needs to solve the differential equations of motion (EoM) on the cosmological background in order to establish whether the model can reproduce the desired, e.g. phantom crossing, EoS. For example, the EoS of quintessence is determined in terms of the kinetic and potential terms of the scalar field, and one can determine its behavior only after solving for the evolution of the field. Alternatively, one could adopt a designer approach, where the background is chosen a priori and the theory is solved accordingly; see, e.g.~\citep{Song:2006ej} for the case of $f(\mathscr{R})$ theories. However, this procedure generally involves differential equations and allows one to only numerically reconstruct the theory up to some boundary conditions. In our work, we build a framework in which the EoS of DE is determined by the free functions of time in a Lagrangian algebraically. In this way, we can control the EoS without needing any knowledge of the solutions for the fields. 

We focus on scalar-tensor models of gravity and will work within the broad framework of degenerate higher order scalar-tensor (DHOST) theories~\citep{Langlois:2015cwa,BenAchour2016a}, which encompasses but goes beyond Horndeski gravity~\citep{Horndeski1974,Deffayet2011}. The DHOST theories in the unitary gauge are commonly referred to as U-DHOST theories and, as was shown in~\citep{Gao2014,Gao2019c}, they can be reformulated as spatially covariant gravity with the velocity of lapse function (SCG-$L$) theories by means of the Arnowitt--Deser--Misner (ADM) formalism. In the latter, time diffeomorphism invariance is broken by fixing the foliation of spacetime, while spatial diffeomorphism invariance is preserved. In this paper, we will adopt the spatially covariant formulation of U-DHOST,  because the Ostrogradsky ghost problem is automatically avoided (time derivatives appear only up to the first order), while providing a vast theoretical space with any higher order spatial derivatives. It also offers a more transparent connection to the EFT formalism, which will be useful when exploring the implications for cosmological observables. We will construct a class of DE theories for which the EoS is directly determined by the free functions of time from the Lagrangian and thus can be set to any desired evolution history, including the phantom crossing ones, without resorting to the designer approach, nor needing the knowledge of background solutions. We will examine the stability of the cosmological perturbations of the models by checking the ghost and gradient instabilities and address to the strong coupling problem as well.

The paper is organized as follows. In Section \ref{sec_Framework},
we illustrate the relations between U-DHOST theories and SCG-$L$ theories. We construct the model in Section \ref{sec_Model} by identifying the EoS of DE and solving the phantom crossing conditions. In Section \ref{sec_Stabilities_Criteria}, we discuss the ghost and gradient instabilities of the model at the linear perturbation level, as well as the strong coupling problem, and explore the cosmological phenomenology. We conclude in Section \ref{sec_Conclusions}. 

\section{The framework}\label{sec_Framework}
We start from the broad DHOST framework~\citep{Langlois:2015cwa}, which was identified as an extension of Horndeski gravity~\citep{Horndeski1974} still propagating only one scalar and two tensorial DoF in the four-dimensional space-time. In this class of theories, when working in unitary gauge, one notices that the velocity of the lapse function appears leading to the beyond-Horndeski terms. We will see in Section \ref{sec_Model} that the term linear in this velocity will be helpful in constructing our model. In this Section, we will show that U-DHOST theories can be recast as a subclass of SCG-$L$ theories \citep{Gao2019c}, which we will use as the starting point for the construction of our model.

We start from the action of DHOST theories \citep{Langlois:2015cwa} which is constructed with the second derivatives of the scalar field up to the quadratic order:
\begin{equation}
S^{\text{DHOST}}=\int\text{d}^{4}x\sqrt{-g}\left[G\left(\phi,X\right)\mathscr{R}+P\left(\phi,X\right)+\mathcal{B}^{\mu\nu}\phi_{\mu\nu}+\mathcal{C}^{\mu\nu\rho\sigma}\phi_{\mu\nu}\phi_{\rho\sigma}\right],\label{Sec-II_S_U-DHOST}
\end{equation}
where we denote
\begin{equation}
\mathcal{B}^{\mu\nu}\equiv B_{1}g^{\mu\nu}+B_{2}\phi^{\mu}\phi^{\nu},\label{Sec-II_B^munu}
\end{equation}
and
\begin{eqnarray}
\mathcal{C}^{\mu\nu\rho\sigma} & \equiv & \frac{1}{2}C_{1}\left(g^{\mu\rho}g^{\nu\sigma}+g^{\mu\sigma}g^{\nu\rho}\right)+C_{2}g^{\mu\nu}g^{\rho\sigma}+\frac{1}{2}C_{3}\left(\phi^{\mu}\phi^{\nu}g^{\rho\sigma}+\phi^{\rho}\phi^{\sigma}g^{\mu\nu}\right)\nonumber \\
 &  & +\frac{1}{4}C_{4}\left(\phi^{\mu}\phi^{\rho}g^{\nu\sigma}+\phi^{\nu}\phi^{\rho}g^{\mu\sigma}+\phi^{\mu}\phi^{\sigma}g^{\nu\rho}+\phi^{\nu}\phi^{\sigma}g^{\mu\rho}\right)+C_{5}\phi^{\mu}\phi^{\nu}\phi^{\rho}\phi^{\sigma},
\end{eqnarray}
with $\mathscr{R}$ representing the 4-dimensional Ricci scalar. The coefficients
$\left\{ G,P,B_{1},B_{2},C_{1,2,3,4,5}\right\} $ are general functions
of $\phi$ and $X\equiv-\phi_{\mu}\phi^{\mu}/2$ and here $\phi_{\mu}\equiv\nabla_{\mu}\phi$.
Usually, the $B_{2}$ term is not considered in (\ref{Sec-II_B^munu})
since it could be absorbed into the $B_{1}$ and $P$ terms via the following relation which is valid up to a total derivative term
\begin{equation}
\int\text{d}^{4}x\sqrt{-g}B_{1}\left(\phi,X\right)\phi^{\mu}{}_{\mu}\simeq\int\text{d}^{4}x\sqrt{-g}\left(B_{1X}\phi^{\mu}\phi^{\nu}\phi_{\mu\nu}+2XB_{1\phi}\right).\label{Sec-II_ibp_covariant}
\end{equation}
However, for reasons that will become clear later,
we keep both terms in (\ref{Sec-II_B^munu}). In general, the coefficients in
(\ref{Sec-II_S_U-DHOST}) are required to satisfy three degeneracy
conditions derived in~\citep{Langlois:2015cwa,BenAchour2016a} to guarantee the absence of the Ostrogradsky ghost. If one works with a timelike scalar field, the conditions can be reduced to a single one~\citep{DeFelice2018,DeFelice:2021hps}:
\begin{equation}
\left[X\left(C_{1}+3C_{2}\right)-G\right]\left[C_{1}+C_{2}-2X\left(C_{3}+C_{4}\right)+4X^{2}C_{5}\right]=3X\left(C_{2}-XC_{3}-G_{X}\right)^{2},\label{Sec-II_Degeneracy_Condition}
\end{equation}
where $G_{X}=\partial G/\partial X$. In this case, it is convenient to work in the unitary gauge i.e., setting $t=\phi$ as a convention, then the derivatives of the scalar field can be decomposed by
\begin{equation}
\phi_{\mu}=-\frac{1}{N}n_{\mu},\quad\phi_{\mu\nu}=-\frac{1}{N}\left(n_{\mu}n_{\nu}\frac{1}{N}F-2n_{(\mu}a_{\nu)}+K_{\mu\nu}\right),\label{Sec-II_Decomposition_phi}
\end{equation}
where
\begin{equation}
F\equiv\pounds_{\vec{n}}N,\quad a_{\mu}\equiv D_{\mu}\ln N,\quad K_{\mu\nu}\equiv\frac{1}{2}\pounds_{\vec{n}}h_{\mu\nu}.\label{Sec-II,F_a_K}
\end{equation}
Here $N$, $n_{\mu}$ and $h_{\mu\nu}$ are, respectively the lapse function,
normal vector and induced metric defined on the spacelike hypersurfaces
specified by the scalar filed; $D_{\mu}$ is the covariant derivative
compatible with $h_{\mu\nu}$ and $\pounds_{\vec{n}}$ represents
the Lie derivative along $n^{\mu}$ with the normalization of $n^{\mu}n_{\mu}=-1$.
With these conventions, one can use the degeneracy condition (\ref{Sec-II_Degeneracy_Condition}) to remove one of the free coefficients in the action; we will use it to eliminate $C_{5}$:
\begin{equation}
C_{5}=\frac{3\left(2N^{2}C_{2}-C_{3}+2N^{5}G_{N}\right)^{2}}{4\left(C_{1}+3C_{2}-2N^{2}G\right)}-N^{2}\left[N^{2}\left(C_{1}+C_{2}\right)-C_{3}-C_{4}\right].\label{Sec-II_C_5}
\end{equation} 
The theories belonging to the class of actions (\ref{Sec-II_S_U-DHOST}) that satisfy  the condition (\ref{Sec-II_Degeneracy_Condition}), or equivalently (\ref{Sec-II_C_5}), are referred to as U-DHOST theories.

By substituting (\ref{Sec-II_Decomposition_phi}) with $C_{5}$ (\ref{Sec-II_C_5}) back into (\ref{Sec-II_S_U-DHOST}) and after some manipulations, one can show that the U-DHOST action can be recast as
\begin{equation}
S=\int\text{d}t\text{d}x^{3}N\sqrt{h}\left[\alpha F+\beta\left(K+\gamma F\right)+\sigma_{1}\left(K_{ij}K^{ij}-\frac{1}{3}K^{2}\right)+\sigma_{2}\left(K+\gamma F\right)^{2}+\mathcal{V}\right]\,.\label{Sec-II_S}
\end{equation}
In fact, the same action has been previously proposed as a concrete example of the SCG-$L$ theories \citep{Gao2019c} and we will use it to build our desired model in this work. The potential function is given by \footnote{It was shown in \citep{Gao2019c} that the potential function can be further promoted to be an arbitrary potential function $\mathcal{V}\left(N,h_{ij},R_{ij};D_{i}\right)$ without changing the number of DoF.}
\begin{equation}
\mathcal{V}\equiv\rho_{0}+\rho_{1}R+\rho_{2}a_{i}a^{i}.
\end{equation}
The coefficients in the action (\ref{Sec-II_S}) are now to be considered general functions of $\left(t,N\right)$ and $R$ represents the 3-dimensional Ricci scalar. The other geometric quantities (\ref{Sec-II,F_a_K}) are expressed in the ADM coordinates as follows
\begin{equation}
F=\frac{1}{N}\left(\partial_{t}N-N^{i}D_{i}N\right),\quad a_{i}=D_{i}\ln N,\quad K_{ij}=\frac{1}{2N}\left(\partial_{t}h_{ij}-D_{i}N_{j}-D_{j}N_{i}\right).\label{Sec-II_F_a_K_unitary}
\end{equation}
From these expressions, it can be seen that $a_{i}$ and $K_{ij}$ are, respectively,
the 3-acceleration and extrinsic curvature of spatial hypersurfaces and $F$ characterizes the velocity of the lapse function. With all the definitions given above, there is an one-to-one map between the two sets of coefficients $\left\{ G,P,B_{1},B_{2},C_{1},C_{2},C_{3},C_{4}\right\} $
and $\left\{ \alpha,\beta,\gamma,\sigma_{1},\sigma_{2},\rho_{0},\rho_{1},\rho_{2}\right\} $, which we report in Appendix \ref{App_UDHOST_SCGL}. This is the reason why we prefer not to remove $B_2$ from action (\ref{Sec-II_B^munu}).
The equivalent expression of eq. (\ref{Sec-II_ibp_covariant}) relating $B_1$ and $B_2$ in the unitary gauge, is a relation between $F$ and $K$ given as follows
\begin{equation}
\int\text{d}t\text{d}^{3}xN\sqrt{h}\beta\left(t,N\right)K\simeq-\int\text{d}t\text{d}^{3}xN\sqrt{h}\left(\partial_{N}\beta F+\frac{1}{N}\partial_{t}\beta\right).\label{Sec-II_K+F}
\end{equation}
As we will see in the next section, keeping the terms linear in $F$ and $K$ in the action will be useful for the construction of the desired DE model.

\section{The Model \label{sec_Model}}

To develop a general DE model that can reproduce any desired expansion history—including scenarios involving phantom divide crossing—we start by deriving the expression for the EoS of DE from the action (\ref{Sec-II_S}). This involves first identifying the energy density and pressure of the DE component on cosmological background.

\subsection{Equation of state of dark energy \label{subsec_EoS_DE}}

We specialize in the Friedmann--Lemaître--Robertson--Walker (FLRW) background 
\begin{equation}
\text{d}s^{2}=-\bar{N}^{2}\text{d}t^{2}+a^{2}\delta_{ij}\text{d}x^{i}\text{d}x^{j},\label{Sec-III-A_FLRW}
\end{equation}
where $\bar{N}\left(t\right)$ is the background value of the lapse
function and $a\left(t\right)$ is the scale factor. In this case, action (\ref{Sec-II_S})
reduces to 
\begin{equation}
S_{\text{0}}=\int\text{d}t\,\text{d}^{3}x\bar{N}a^{3}\left[\sigma_{2}\left(3H+\gamma\bar{F}\right)^{2}+\beta\left(3H+\gamma\bar{F}\right)+\alpha\bar{F}+\rho_{0}\right],\label{Sec-III-A_S_bg}
\end{equation}
where we define $H\equiv\dot{a}/a$ and $\bar{F}\equiv\dot{\bar{N}}$
with the ``$\cdot$'' denoting  $\frac{1}{\bar{N}}\frac{\partial}{\partial t}$
and the coefficients are understood as functions of $\left(t,\bar{N}\right)$.
By varying the background action with respect to $\bar{N}$ and $a$ respectively, we can derive the background EoMs. Since we do not include matter sectors other than dark energy in our action, we use the EoMs to determine the DE energy density and pressure as follows:
\begin{equation}
6H^{2}=\rho_{\text{de}}\left(\dot{H},\dot{\bar{F}},H,\bar{F},\bar{N}\right),\quad-4\dot{H}-6H^{2}=p_{\text{de}}\left(\dot{H},\dot{\bar{F}},H,\bar{F},\bar{N}\right),\label{Sec-III-A_EoM_bg}
\end{equation}
where we set $M_\text{Pl}^2/2=1$ for the Planck mass and the RHS of eq. (\ref{Sec-III-A_EoM_bg}) are given by
\begin{equation}
\rho_{\text{de}}=\mathcal{W}_{\dot{H}}\dot{H}+\mathcal{W}_{\dot{F}}\dot{\bar{F}}+\mathcal{W}_{H^{2}}H^{2}+\mathcal{W}_{HF}H\bar{F}+\mathcal{W}_{F^{2}}\bar{F}^{2}+\mathcal{W}_{H}H+\mathcal{W}_{F}\bar{F}+\mathcal{W}_{N},\label{Sec-III-A_rho_de}
\end{equation}
with
\begin{equation}
p_{\text{de}}=\mathcal{M}_{\dot{H}}\dot{H}+\mathcal{M}_{\dot{F}}\dot{\bar{F}}+\mathcal{M}_{H^{2}}H^{2}+\mathcal{M}_{HF}H\bar{F}+\mathcal{M}_{F^{2}}\bar{F}^{2}+\mathcal{M}_{H}H+\mathcal{M}_{F}\bar{F}+\mathcal{M}_{N}.\label{Sec-III-A_p_de}
\end{equation}
The concrete expressions of the above coefficients are reported in Appendix \ref{App_energy_pressure_DE}. With these, we can then define the DE EoS as the following
\begin{equation}
w_{\text{de}}\left(t\right)\equiv\frac{p_{\text{de}}\left(\dot{H},\dot{\bar{F}},H,\bar{F},\bar{N}\right)}{\rho_{\text{de}}\left(\dot{H},\dot{\bar{F}},H,\bar{F},\bar{N}\right)},\label{Sec-III-A_w_de}
\end{equation}
where $H\left(t\right)$ and $\bar{N}\left(t\right)$ should be understood as the solutions of the background EoMs (\ref{Sec-III-A_rho_de}) and (\ref{Sec-III-A_p_de}).
The latter can generally be found via a proper ans\"atze for the particular solutions or numerically by setting appropriate initial conditions. However in this work, we would like to have a more direct control on $w_{\text{de}}$, which does not require one to first solve the EoMs. We address this in the next section. 

\subsection{Phantom crossing condition \label{subsec_PC_condition}}

If we want our DE model to cross the phantom divide at a given time $t_{\ast}$, then we need to impose the following condition on the EoS (\ref{Sec-III-A_w_de}) 
\begin{equation}
w_{\text{de}}\left(t_{\ast}\right)+1=0.\label{Sec-III-B_w_de+1=00003D0}
\end{equation}
Now we insert the expressions for $\rho_{\text{de}}$ and $p_{\text{de}}$ from (\ref{Sec-III-A_rho_de}) and (\ref{Sec-III-A_p_de}) respectively, and find
\begin{equation}
w_{\text{de}}\left(t\right)+1=\frac{\mathcal{D}_{\dot{H}}\dot{H}+\mathcal{D}_{\dot{F}}\dot{\bar{F}}+\mathcal{D}_{H^{2}}H^{2}+\mathcal{D}_{HF}H\bar{F}+\mathcal{D}_{F^{2}}\bar{F}^{2}+\mathcal{D}_{H}H+\mathcal{D}_{F}\bar{F}+\mathcal{D}_{N}}{\mathcal{W}_{\dot{H}}\dot{H}+\mathcal{W}_{\dot{F}}\dot{\bar{F}}+\mathcal{W}_{H^{2}}H^{2}+\mathcal{W}_{HF}H\bar{F}+\mathcal{W}_{F^{2}}\bar{F}^{2}+\mathcal{W}_{H}H+\mathcal{W}_{F}\bar{F}+\mathcal{W}_{N}},\label{Sec-III-B_w_de+1}
\end{equation}
where  
\begin{eqnarray}
\mathcal{D}_{\dot{H}} & \equiv & 2\left[3\sigma_{2}\left(\bar{N}\gamma-1\right)-2\right],\quad\mathcal{D}_{\dot{F}}\equiv2\sigma_{2}\gamma\left(\bar{N}\gamma-1\right),\\
\mathcal{D}_{H^{2}} & \equiv & 18\bar{N}^{2}\sigma_{2}\left(\gamma-\frac{1}{2}\partial_{N}\ln\sigma_{2}\right),\quad\mathcal{D}_{HF}\equiv6\bar{N}\sigma_{2}\left[\beta\left(\bar{N}\gamma+1\right)-\partial_{N}\ln\sigma_{2}\right],\\
\mathcal{D}_{F^{2}} & \equiv & 2\bar{N}\gamma\sigma_{2}\left[\left(\frac{1}{2}\bar{N}\gamma-1\right)\partial_{N}\ln\sigma_{2}+\left(\bar{N}\gamma-1\right)\partial_{N}\ln\gamma+\gamma\right],\\
\mathcal{D}_{H} & \equiv & 6\bar{N}\left[\frac{1}{2}\bar{N}\left(\alpha+\beta\gamma-\partial_{N}\beta\right)+\partial_{t}\left(\gamma\sigma_{2}\right)-\frac{1}{\bar{N}}\partial_{t}\sigma_{2}\right],\\
\mathcal{D}_{F} & \equiv & 2\bar{N}\left[\frac{1}{2}\left(\alpha-\partial_{N}\beta+\beta\gamma\right)+\gamma\left(\gamma\partial_{t}\sigma_{2}+2\sigma_{2}\partial_{t}\gamma\right)-\frac{1}{\bar{N}}\partial_{t}\left(\gamma\sigma_{2}\right)\right],\\
\mathcal{D}_{N} & \equiv & \bar{N}\left[\partial_{t}\alpha+\partial_{t}\left(\beta\gamma\right)-\bar{N}\partial_{N}\rho_{0}\right]-\partial_{t}\beta.\label{Sec-III-B_D_N}
\end{eqnarray}
We aim to determine the condition for phantom crossing in complete generality—that is, to derive a constraint on the functions in our action that ensures phantom crossing occurs regardless of the specific dynamics and solutions of the background fields. One approach to achieve this is to demand that eq. (\ref{Sec-III-B_w_de+1=00003D0}) takes the form of an algebraic equation in $\bar{N}$ rather than a differential one, i.e. we set
\begin{equation}
\mathcal{D}_{\dot{F}}=\mathcal{D}_{\dot{H}}=\mathcal{D}_{H^{2}}=\mathcal{D}_{HF}=\mathcal{D}_{F^{2}}=\mathcal{D}_{H}=\mathcal{D}_{F}=0.\label{Sec-III-B_D_eqs}
\end{equation}
In this way, the only term left in $w_{\textrm de}$ (\ref{Sec-III-B_w_de+1})
is $\mathcal{D}_{N}$ (\ref{Sec-III-B_D_N}) which depends on $\bar{N}$
algebraically. At this stage, it seems like we may still need to know the solution
of $\bar{N}$. However, we will show later that this can
be avoided by introducing a further requirement on the coefficient $\rho_{0}$. Before doing that, let us solve the conditions in eq. (\ref{Sec-III-B_D_eqs}).

\subsubsection{Solving the conditions}
We start from the first two conditions,  $\mathcal{D}_{\dot{F}}=\mathcal{D}_{\dot{H}}=0$, which correspond to 
\begin{equation}
\sigma_{2}\left(\bar{N}\gamma-1\right)\gamma=0,\quad3\sigma_{2}\left(\bar{N}\gamma-1\right)-2=0.
\end{equation}
One can easily notice that the only possible solutions to these equations are
\begin{equation}
\gamma=0,\quad\sigma_{2}=-\frac{2}{3},\label{Sec-III-B_gamma}
\end{equation}
which incidentally also satisfy the next three conditions, $\mathcal{D}_{H^{2}}=\mathcal{D}_{HF}=\mathcal{D}_{F^{2}}=0$. 
The remaining equations $\mathcal{D}_{H}=\mathcal{D}_{F}=0$ lead to the same condition in the coefficients, i.e.: 
\begin{equation}
\bar{N}\left(\alpha-\partial_{N}\beta\right)=0, \quad\Rightarrow\alpha=\frac{\partial\beta}{\partial N}.\label{Sec-III-B_alpha}
\end{equation}
One may be concerned about the denominator of eq. (\ref{Sec-III-B_w_de+1}) (i.e. $\rho_\text{de}$ defined in (\ref{Sec-III-A_rho_de})) vanishes with solutions (\ref{Sec-III-B_gamma}) and (\ref{Sec-III-B_alpha}). We show that with these solutions, the $\rho_\text{de}$ (\ref{Sec-III-A_rho_de}) and $p_\text{de}$ (\ref{Sec-III-A_p_de}) are simplified to  
\begin{equation}
\rho_{\text{de}}=-\partial_{N}\left[\bar{N}\left(\rho_{0}-\frac{1}{\bar{N}}\partial_{t}\beta\right)\right],\quad p_{\text{de}}=\rho_{0}-\frac{1}{\bar{N}}\partial_{t}\beta,
\end{equation}
and thus they do not vanish in general. 

\subsubsection{Freezing Gravity}

With the solutions derived above, the equations (\ref{Sec-III-B_D_eqs}) are all satisfied and the phantom crossing condition (\ref{Sec-III-B_w_de+1=00003D0}) reduces to 
\begin{equation}
\mathcal{D}_{N}\left(t_{\ast}\right)=\left.-\bar{N}^{2}\partial_{N}\left(\rho_{0}-\frac{1}{\bar{N}}\partial_{t}\beta\right)\right|_{t=t_{\ast}}=0,\label{Sec-III-B_D_N-1}
\end{equation} 
and the background action (\ref{Sec-III-A_S_bg}) is simplified as
\begin{equation}
S_{\text{0}}\simeq\int\text{d}t\text{d}^{3}x\bar{N}a^{3}\left(-6H^{2}+\rho_{0}-\frac{1}{\bar{N}}\partial_{t}\beta\right).\label{Sec-III-B_S_0_1}
\end{equation}
In order to solve (\ref{Sec-III-B_D_N-1}), we need to know $\bar{N}$ as a solution to the EoMs derived from the background action (\ref{Sec-III-B_S_0_1}), which is something we would like to avoid. We can address this issue by restoring the symmetry of time reparametrization that is broken in the action (\ref{Sec-III-B_S_0_1}). By reinstating time reparametrization invariance, we gain the freedom to choose the background value of the lapse to fix the time coordinate and it thus becomes not physically meaningful. This can be achieved by making the following choice
\begin{equation} 
\rho_{0}\left(t,\bar{N}\right)=\frac{1}{\bar{N}}\partial_{t}\beta-\frac{4}{\bar{N}}\partial_{t}\mathcal{F}-6\mathcal{G}^{2},\label{Sec-III-B_rho_0}
\end{equation}
where $\mathcal{G}$ and $\mathcal{F}$ are introduced as two generic functions of time and with this choice as well as taking an integration by parts, the background action (\ref{Sec-III-B_S_0_1}) becomes
\begin{equation}
S_{\text{0}}\simeq-6\int\text{d}t\text{d}^{3}x\bar{N}a^{3}\left(H^{2}-2\mathcal{F}H+\mathcal{G}^{2}\right).\label{Sec-III-B_S_0_2}
\end{equation}
Recalling that the Hubble parameter was defined through the time-reparametrization invariant derivative, we notice that action (\ref{Sec-III-B_S_0_2}) is now invariant under time reparametrization and we thus have freedom to set $\bar{N}=1$, simplifying the EoMs to
\begin{equation}
H^{2}=\mathcal{G}^{2},\quad\dot{H}=\dot{\mathcal{F}}\,.\label{Sec-III-B_beta=00003D0_EoM_BG}
\end{equation}
The energy density and pressure of DE then are read as follows
\begin{equation}
\rho_{\text{de}}\left(t\right)=6\mathcal{G}^{2},\quad p_{\text{de}}\left(t\right)=-6\mathcal{G}^{2}-4\dot{\mathcal{F}},\label{Sec-III-B_rho_de_p_de}
\end{equation}
and the EoS and phantom crossing condition are simply given by
\begin{equation}
w_{\text{de}}\left(t\right)=\frac{-3\mathcal{G}^{2}-2\dot{\mathcal{F}}}{3\mathcal{G}^{2}},\quad w_{\text{de}}\left(t_{\ast}\right)+1=\left.-\frac{2}{3}\frac{\dot{\mathcal{F}}}{\mathcal{G}^{2}}\right|_{t=t_{\ast}}=0.\label{Sec-III_w_de}
\end{equation}
From which we see that on the cosmological background (\ref{Sec-III-A_FLRW}),
only $\mathcal{F}\left(t\right)$ and $\mathcal{G}\left(t\right)$ enter into the action. From the corresponding Friedmann equations (\ref{Sec-III-B_beta=00003D0_EoM_BG}), we find the general solutions $H=\pm\mathcal{G}$, and we fix the sign by requiring $H>0$. The other function, 
$\mathcal{F}$, shall agree with $\mathcal{G}$ up to an integration
constant. The background evolution, and DE EoS, are thus completely determined by the given $\mathcal{G}$ and $\mathcal{F}$. Additionally, using the second equation in (\ref{Sec-III_w_de}) one can identify $\mathcal{G}$ and $\mathcal{F}$ such that there is a phantom crossing at any given $t_\ast$.

To summarize, we have thus built the following general DE model 
\begin{equation}
S=\int\text{d}t\text{d}^{3}xN\sqrt{h}\left[\sigma_{1}\left(K_{ij}K^{ij}-\frac{1}{3}K^{2}\right)+\rho_{1}R+\rho_{2}a_{i}a^{i}-\frac{2}{3}K^{2}+4\mathcal{F}K-6\mathcal{G}^{2}\right],\label{Sec-III-B_S_PCDE}
\end{equation}
where $\sigma_{1}$, $\rho_{1}$ and $\rho_{2}$ are arbitrary functions
of $\left(t,N\right)$ and $\mathcal{F}$ and $\mathcal{G}$ are two free functions of time only. One can use them to reproduce any given DE EoS and, if desired, set a phantom crossing at a given time $t_\ast$. We have thus accomplished our goal of constructing an action for a DE model that enables direct control over the DE equation of state, independently of the dynamics of the background fields. Some comments are in order.

A crucial element in achieving this was the inclusion of the term linear in the extrinsic curvature, $4\mathcal{F}K$ in the action. From (\ref{Sec-II_K+F}) one can see that $\mathcal{F}K\simeq-\partial_{t}\mathcal{F}/N$ which can be further  recast as  $\mathcal{F}_{\phi}\sqrt{2X}$ after we recover general covariance. The latter operator was first considered in the \emph{cuscuton} theory of~\citep{Afshordi2007,Afshordi2007a}, where the scalar field has an infinite speed of sound, thus it does not represent an independent DoF but rather propagates along
with metric fields. When the cuscuton theory is formulated in unitary gauge, the scalar mode is completely suppressed and the theory is considered a minimal modification of the cosmological constant. This idea was later generalized in \citep{Cooney:2009aa,DeFelice2020b,Iyonaga2020,Aoki2020b,Felice:2022ab,Ganz2022,Bazeia:2025aa} (see also \citep{Lin:2017oow,Iyonaga2018,Gao:2019twq}) where it was shown that a non-propagating scalar DoF can address dark energy problem. In our model, the scalar DoF is generally propagating unless the coefficients $\sigma_{1}$, $\rho_{1}$ and $\rho_{2}$ are chosen particularly \citep{Gao:2019twq}. Nevertheless, we will see in the next section that the sound speed of the scalar mode approaches infinity on large scales, thus the mode tends to become non-propagating in that limit. Given this asymptotic behavior, one may be concerned about the strong coupling problem \citep{Deffayet:2005aa}. However, as we will show in the next section, this asymptotic behavior persists at least up to cubic order. This motivates us to speculate that the scalar DoF remains frozen at arbitrary perturbative order in the infrared (IR) limit, potentially avoiding the associated strong-coupling issue. Due to this characteristic IR behavior, hereafter we dub the model (\ref{Sec-III-B_S_PCDE}) \emph{Freezing Gravity}.

In Appendix \ref{App_UDHOST_SCGL} we provide the mapping between U-DHOST and SCG-$L$ which can be used to straightforwardly derive the scalar-tensor formulation of Freezing Gravity.

\section{Stability of linear perturbations\label{sec_Stabilities_Criteria}}
We shall now examine the stability of Freezing Gravity in the linear cosmological perturbations. After identifying the necessary conditions to ensure that ghost and gradient instabilities are avoided, we will explore the predictions of Freezing Gravity for some cosmological observables. For this part of the work, it will be convenient to work with the language of EFT of DE, which will also allow us to directly use the Einstein-Boltzmann solver \texttt{EFTCAMB}.

\subsection{Quadratic action of EFT}
We are interested in the stability properties of our model, as well as in the dynamics of linear perturbations. We shall therefore focus on the quadratic action of Freezing Gravity (\ref{Sec-III-B_S_PCDE}), and map it into the formalism of EFT of DE. The resulting action reads
\citep{Gubitosi:2012hu,Gleyzes:2014rba,Langlois2017a}  
\begin{eqnarray}
S_{\text{2}}^{\text{EFT}} & = & \int\text{d}t\,\text{d}x^{3}a^{3}\frac{M^{2}}{2}\Big[\delta K^{i}{}_{j}\delta K^{j}{}_{i}-\left(1+\frac{2}{3}\alpha_{\text{L}}\right)\delta K^{2}+\left(1+\alpha_{\text{H}}\right)\delta R\delta N\nonumber \\
 &  & \qquad\qquad\qquad\quad+\left(1+\alpha_{\text{T}}\right)\left(\delta_{1}R\frac{\delta\sqrt{h}}{a^{3}}+\delta_{2}R\right)+\frac{\beta_{3}}{a^{2}}\left(\partial_{i}\delta N\right)^{2}\Big],\label{Sec-IV-A_quadrtic_action}
\end{eqnarray}
where $M$ is the effective Planck mass with the dimensionless parameters given by
\begin{equation}
\frac{M^{2}}{2}=\sigma_{1},\quad\alpha_{\text{L}}=\frac{1}{\sigma_{1}}-1,\quad\alpha_{\text{T}}=\frac{\rho_{1}}{\sigma_{1}}-1,\quad\alpha_{\text{H}}=\frac{\rho_{1}+\partial_{N}\rho_{1}}{\sigma_{1}}-1,\quad\beta_{3}=\frac{\rho_{2}}{\sigma_{1}},\label{Sec-IV-A_EFT_params}
\end{equation}
again we have set $M_{\text{Pl}}^{2}/2=1$ for the Planck mass. One can notice that the effective Planck mass, $M$, and $\alpha_{\text{L}}$,
which represents the detuning of the kinetic terms of general relativity,
are not independent within Freezing Gravity. In particular, imposing $M^{2}>0$ implies $\alpha_{\text{L}}>-1$ which will be useful in discussing stable regimes in the next sections. For convenience, we also give the EFT action in terms of the mass parameters $M_{i}$ in Appendix \ref{App_M_functions}, while we will keep using the $\alpha_{i}$ basis in the following discussions. 

\subsection{Effective action for linear perturbations}
In order to analyze the stability of Freezing Gravity, we derive the effective quadratic action in Fourier space, after integrating out the auxiliary
perturbations. We focus on the scalar and tensor perturbations denoted by $\zeta$ and ${\gamma}_{ij}$ respectively and find
\begin{equation}
S_{2}^{\text{eff}}=\int\text{d}t\frac{\text{d}^{3}k}{\left(2\pi\right)^{3}}a^{3}\frac{M^{2}}{2}\left[\mathcal{A}_{\dot{\zeta}^{2}}\left(\dot{\zeta}^{2}-c_{s}^{2}\frac{k^{2}}{a^{2}}\zeta^{2}\right)+\frac{1}{4}\left(\dot{\gamma}_{ij}^{2}-c_{\text{T}}^{2}\frac{k^{2}}{a^{2}}\gamma_{ij}^{2}\right)\right],\label{Sec-IV-B_S_eff}
\end{equation}
where the sound speeds of the scalar and tensor modes are defined by 
\begin{equation}
c_{s}^{2}\equiv\frac{\mathcal{B}_{\zeta^{2}}}{\mathcal{A}_{\dot{\zeta}^{2}}},\quad c_{\text{T}}^{2}\equiv1+\alpha_{\text{T}},\label{Sec-IV_sound_speed}
\end{equation}
and the coefficients in the above are given as follows 
\begin{equation}
\mathcal{A}_{\dot{\zeta}^{2}}\equiv\frac{6\left(\alpha_{\text{L}}+1\right)\beta_{3}\frac{k^{2}}{a^{2}}}{6H^{2}\left(\alpha_{\text{L}}+1\right)+\alpha_{\text{L}}\beta_{3}\frac{k^{2}}{a^{2}}},\quad\mathcal{B}_{\zeta^{2}}\equiv\frac{\mathcal{B}_{0}+\mathcal{B}_{2}\frac{k^{2}}{a^{2}}+\mathcal{B}_{4}\frac{k^{4}}{a^{4}}}{\left[6H^{2}\left(\alpha_{\text{L}}+1\right)+\alpha_{\text{L}}\beta_{3}\frac{k^{2}}{a^{2}}\right]^{2}}.\label{Sec-IV-B_A=000026B}
\end{equation}
Due to the length, we report the expressions of $\mathcal{B}_{0}$, $\mathcal{B}_{2}$ and
$\mathcal{B}_{4}$ in terms of $\left\{ \alpha_{\text{T}},\alpha_{\text{L}},\alpha_{\text{H}},\beta_{3}\right\} $ in Appendix \ref{App_gradient_coeff}.
The ghost and gradient stabilities of the scalar and tensor perturbations require the following conditions respectively
\begin{equation}
\mathcal{A}_{\dot{\zeta}^{2}}\geq0,\quad\mathcal{B}_{\zeta^{2}}\geq0,\quad M^{2}\geq0,\quad\alpha_{\text{T}}\geq-1.\label{Sec-IV_stability_conds}
\end{equation}
Before deriving the valid regimes of the EFT parameters satisfying these criteria, we can already get some indications by looking into the long- and short-wavelength limits of eq. (\ref{Sec-IV-B_A=000026B}). For simplicity, we will consider the case of luminal speed of gravitational waves i.e., $\alpha_{\text{T}}=0$, in the following discussion.
\begin{itemize}
\item In the long-wavelength limit, we have
\begin{equation}
\lim_{k\rightarrow0}\mathcal{A}_{\dot{\zeta}^{2}}=0,\quad\lim_{k\rightarrow0}\mathcal{B}_{\zeta^{2}}=2\left(\alpha_{\text{H}}+1\right)\left[\frac{\alpha_{\text{H}}}{\alpha_{\text{H}}+1}-\frac{\dot{H}}{H^{2}}+\frac{1}{H}\frac{\text{d}}{\text{d}t}\ln\left(\frac{\alpha_{\text{H}}+1}{\alpha_{\text{L}}+1}\right)\right].\label{Sec-IV-B_IR}
\end{equation}
Since $\mathcal{A}_{\dot{\zeta}^{2}}$ vanishes  the ghost stability criterion is trivially satisfied in this limit. However, a vanishing kinetic term may signal a strong coupling problem. We argue that this issue is circumvented because the dynamic terms of the higher order effective action vanish at every order in the IR limit. One can check that this property holds at cubic order by looking into the relevant terms of the cubic Lagrangian
\begin{eqnarray}
\mathcal{L}_{3}^{\text{kin}} & \supset & \mathcal{C}_{\zeta\dot{\zeta}^{2}}\zeta\dot{\zeta}^{2}+\mathcal{C}_{\dot{\zeta}^{2}\delta N}\dot{\zeta}^{2}\delta N+\mathcal{C}_{\zeta\dot{\zeta}\delta N}\zeta\dot{\zeta}\delta N+\mathcal{C}_{\zeta^{2}\delta N}\zeta^{2}\delta N\nonumber \\
 &  & +\mathcal{C}_{\dot{\zeta}\delta N^{2}}\dot{\zeta}\left(\delta N\right)^{2}+\mathcal{C}_{\zeta\delta N^{2}}\zeta\left(\delta N\right)^{2}+\mathcal{C}_{\delta N^{3}}\left(\delta N\right)^{3}\,.
\end{eqnarray}
After integrating out the auxiliary field $\delta N$, one finds
\begin{equation}
\mathcal{L}_{3}^{\text{eff}}\left[\zeta,\dot{\zeta}\right]\supset\frac{\tilde{\mathcal{C}}_{\dot{\zeta}^{3}}\dot{\zeta}^{3}+\left(\tilde{\mathcal{C}}_{\zeta\dot{\zeta}^{2}}^{\left(2\right)}+\tilde{\mathcal{C}}_{\zeta\dot{\zeta}^{2}}^{\left(4\right)}\frac{k^{2}}{a^{2}}\right)\zeta\dot{\zeta}^{2}+\tilde{\mathcal{C}}_{\dot{\zeta}\zeta^{2}}\frac{k^{2}}{a^{2}}\zeta^{2}\dot{\zeta}}{\left[6H^{2}\left(\alpha_{\text{L}}+1\right)+\alpha_{\text{L}}\beta_{3}\frac{k^{2}}{a^{2}}\right]^{2}}\frac{k^{2}}{a^{2}}\,.\label{Sec-IV-B_L_3}
\end{equation}
The coefficients with tilde are independent of $k$ thus the
dynamical terms vanish in the $k\rightarrow0$ limit. We speculate that this will happen at every order suggesting that the apparent strong-coupling issue may be avoided, with the scalar mode turning into the generalized instantaneous/shadowy mode \citep{DeFelice2018,DeFelice:2021hps} and being frozen in the large scale limit. We point out that it is generally difficult to prove this order by order, and one possible way to show it is to perform a Hamiltonian analysis of Freezing Gravity which counts the number of DoF non-perturbatively. In this way, we may be able to show that the number of DoF reduces from three to two in the large scale limit so that the scalar mode is frozen non-perturbatively. However, a detailed Hamiltonian analysis is beyond the scope of this work and we leave this for future work. A similar argument holds also for the case of $\beta_{3}=0$, in which the kinetic term $\mathcal{A}_{\dot{\zeta}^{2}}$ vanishes identically. Without loss of generality, we consider the case of $\beta_{3}\neq0$ in the following discussions. 
\item In the short-wavelength limit, we have 
\begin{equation}
\lim_{k\rightarrow\infty}\mathcal{A}_{\dot{\zeta}^{2}}=6\left(1+\frac{1}{\alpha_{\text{L}}}\right),\quad\lim_{k\rightarrow\infty}\mathcal{B}_{\zeta^{2}}=2\left[\frac{2}{\beta_{3}}\left(\alpha_{\text{H}}+1\right)^{2}-1\right],\label{Sec-IV-B_UV}
\end{equation}
from which we see that $\mathcal{B}_{\zeta^{2}}\geq0$ gives $\beta_{3}>0$ as a necessary but not sufficient condition and the sign of $\mathcal{A}_{\dot{\zeta}^{2}}$ changes according to the ranges of $\alpha_{\text{L}}$. From eq. (\ref{Sec-IV-A_EFT_params}) we see that a positive $M^2$ requires $\alpha_{\text{L}}\in\left(-1,+\infty\right)$. One may think that $-1<\alpha_{\text{L}}<0$ should be excluded since $\mathcal{A}_{\dot{\zeta}^{2}}<0$ in the ultraviolet (UV) limit. However, as we will see in the following derivations, in this case there is a pole in the sound speed of the scalar mode (\ref{Sec-IV_sound_speed}) which gives a cut-off energy scale and the perturbations can be maintained stable under this scale. Therefore, we will discuss the stability criteria separately for the cases of $-1<\alpha_{\text{L}}<0$ and $\alpha_{\text{L}}\geq0$. 
\end{itemize}

\subsection{Stability criteria}
We shall now derive the exact criteria to avoid ghost and gradient instabilities. Following the asymptotic analysis, we will treat separately the $-1<\alpha_{\text{L}}<0$ and $\alpha_{\text{L}}\protect\geq0$ cases.

\subsubsection{$-1<\alpha_{\text{L}}<0$ case}
In the  $-1<\alpha_{\text{L}}<0$ case,  with $\beta_{3}>0$, there
is a pole in the speed of sound (\ref{Sec-IV_sound_speed}) with the coefficient (\ref{Sec-IV-B_A=000026B}) giving the following energy cut-off
\begin{equation}
k_{\star}^{2}=-6\frac{\alpha_{\text{L}}+1}{\alpha_{\text{L}}\beta_{3}}\left(aH\right)^{2}.\label{Sec-IV-C_cut-off_scale}
\end{equation}
Beyond this scale, higher order terms  should be taken into account. 
In order for the theory to be valid over linear cosmological scales, we require the cut-off to be much smaller than the Hubble scale, i.e. we set the cut-off scale to be $k_{\star}/(aH)\equiv\lambda \gg 1$. This can be translated into an upper bound to $\beta_{3}$ 
\begin{equation}
\beta_{3}<-\frac{6}{\lambda^2}\left(1+\frac{1}{\alpha_{\text{L}}}\right),\label{Sec-IV-C_beta_3_ub}
\end{equation}
and the ghost instability is then avoided below the cut-off scale 
$k_{\star}$. Furthermore, requiring that  $\mathcal{B}_{0}$, $\mathcal{B}_{2}$ and $\mathcal{B}_{4}$ in (\ref{Sec-IV-B_A=000026B}) are non-negative, we derive the following sufficient conditions for the avoidance of the gradient instability: 
\begin{equation}
-1<\alpha_{\text{L}}<0,\quad0<\beta_{3}<-\frac{6}{\lambda^2}\left(1+\frac{1}{\alpha_{\text{L}}}\right),\quad\frac{\dot{H}}{H^{2}-\dot{H}}\leq\alpha_{\text{H}}\leq-\frac{1}{4},\label{Sec-IV-C_valid_alpha}
\end{equation}
and
\begin{equation}
\frac{\text{d}\alpha_{\text{L}}}{\text{d}t}\leq0,\quad\frac{\text{d}\beta_{3}}{\text{d}t}\geq0,\quad\frac{\text{d}}{\text{d}t}\ln\left|\alpha_L\right|\geq\frac{\text{d}}{\text{d}t}\ln\left(\alpha_{\text{H}}+1\right)\geq\frac{\text{d}}{\text{d}t}\ln\left(\alpha_{\text{L}}+1\right).\label{Sec-IV-C_dalpha/dt}
\end{equation}

\subsubsection{$\alpha_{\text{L}}\geq0$ case}
In the $\alpha_{\text{L}}\geq0$ with $\beta_{3}>0$ case, the
ghost instability is automatically avoided since $\mathcal{A}_{\dot{\zeta}^{2}}$
in eq. (\ref{Sec-IV-B_A=000026B}) is always positive. On the other hand, the gradient instability can be avoided by imposing
\begin{equation}
\alpha_{\text{L}}\geq0,\quad2\left(\alpha_{\text{H}}+1\right)^{2}\geq\beta_{3}>0,\quad\alpha_{\text{H}}\geq\frac{\dot{H}}{H^{2}-\dot{H}},\label{Sec-IV_aL>0}
\end{equation}
with
\begin{equation}
\frac{\text{d}\alpha_{\text{L}}}{\text{d}t}\leq0,\quad\frac{\text{d}\beta_{3}}{\text{d}t}\leq0,\quad\frac{\text{d}\alpha_{\text{H}}}{\text{d}t}\geq0.\label{Sec-IV_daL/dt<0}
\end{equation}
 As an example, it can be easily seen that these conditions do in fact correspond to a positive $\quad\mathcal{B}_{\zeta^{2}}$, by looking at the $\alpha_{\text{L}}=0$ case, where  the coefficients (\ref{Sec-IV-B_A=000026B}) drastically simplify to 
\begin{equation}
\mathcal{A}_{\dot{\zeta}^{2}}=\left(\frac{k}{aH}\right)^{2}\beta_{3},\quad\mathcal{B}_{\zeta^{2}}=2\left(\alpha_{\text{H}}+1\right)\left[\frac{\alpha_{\text{H}}}{\alpha_{\text{H}}+1}-\frac{\dot{H}}{H^{2}}+\frac{1}{H}\frac{\text{d}}{\text{d}t}\ln\left(\alpha_{\text{H}}+1\right)\right],\label{Sec-IV_aL=00003D0_A=000026B}
\end{equation}
and it is easy to see that they are positive with the conditions (\ref{Sec-IV_aL>0}) and (\ref{Sec-IV_daL/dt<0}). 

\begin{figure}[H]
\begin{centering}
\includegraphics[scale=0.50]{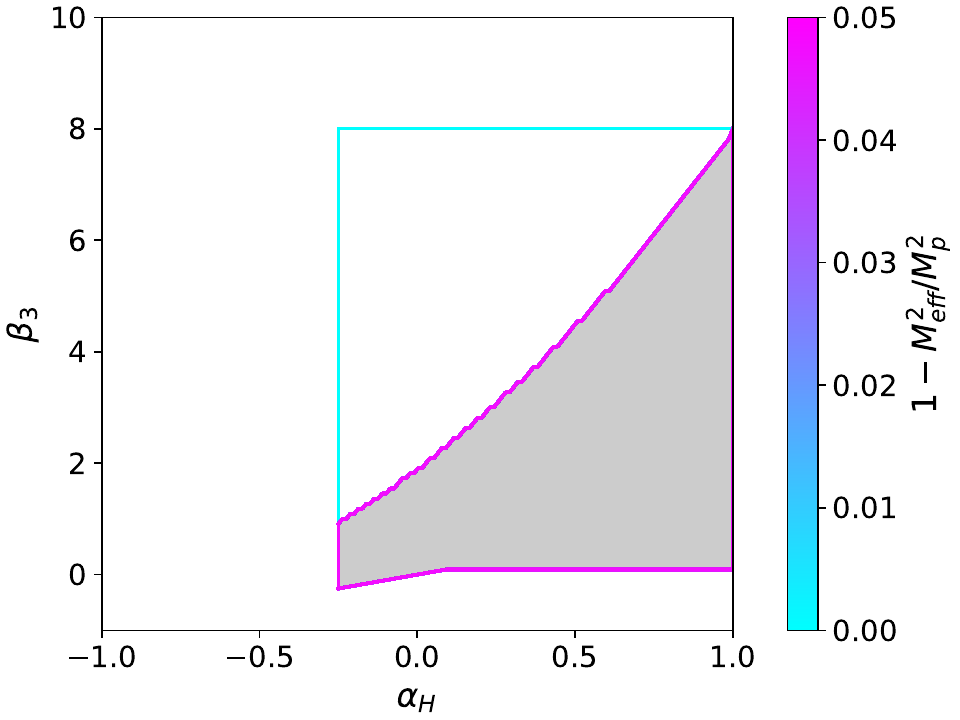}
\par\end{centering}
\caption{Stable regions for $\alpha_{\text{H}}$ and $\beta_{3}$ are drawn
with the deviations of the Planck mass square up to 5\%.  $\alpha_{\text{H}}$
and $\beta_{3}$ are constrained within the gray region when $M^{2}<M_{\text{Pl}}^{2}$
which agree with the bounds given in eq. (\ref{Sec-IV_aL>0}). In
particular when $M^{2}=M_{\text{Pl}}^{2}$, $\beta_{3}$ is not bounded
by $\alpha_{\text{H}}$ therefore the valid values are taken on the cyan frame. The values of $k$ explored to study the stability are in the range $10^{-4}\sim
1 \ \rm{Mpc}^{-1}$.} \label{Fig_valid_regimes}
\end{figure}

In order to visualize the valid regimes of this case, we draw some plots of the stable parameter space by means of the stability module in $\mathtt{EFTCAMB}$ \citep{Hu:2014aa,Raveri:2014aa}. For this, we restrict ourselves to the case of constant EFT parameters. In Fig. \ref{Fig_valid_regimes}, the plot is drawn with the
effective Planck mass deviating from the Planck mass up to $5\%$ and the valid regimes of $\alpha_{\text{H}}$ and $\beta_{3}$ are constrained within the shaded area, since the $\mathcal{B}_{4}$ term in (\ref{Sec-IV-B_A=000026B}) quickly dominates in the small scale if $M^{2}<M_{\text{Pl}}^{2}$. In particular, if there is no deviation from the Planck mass, i.e. $\alpha_{\text{L}}=0$, the values of $\alpha_{\text{H}}$ and $\beta_{3}$ can be taken independently on the cyan frame because the $\mathcal{B}_{4}$ and $\mathcal{B}_{2}$ terms (see in eq. (\ref{App_B_B_4}) and (\ref{App_B_B_2})) vanish and only $\mathcal{B}_{0}$ term (\ref{App_B_B_0}) takes effect in
this case \footnote{In order to see gradual change from the cyan frame to the shaded area, one would need a much smaller increment in $1-M^{2}/M_{\text{Pl}}^{2}$ i.e., $\Delta M^2/M_{\text{Pl}}^{2}<10^{-4}$.}.

\begin{figure}[H]
\begin{centering}
\includegraphics[width=\textwidth]{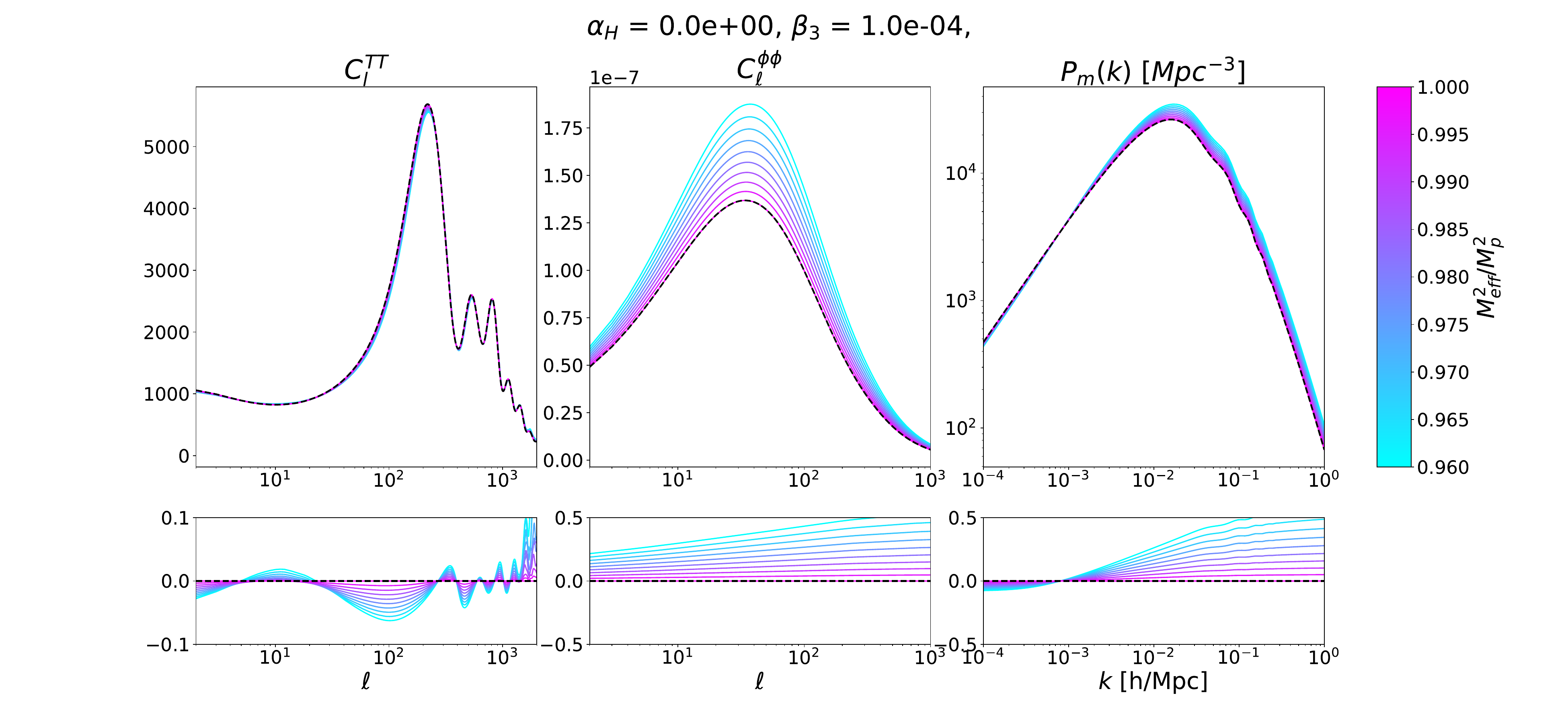}
\par\end{centering}
\caption{Temperature and weak lensing angular power spectra of CMB, $C_{l}^{\text{TT}}$
and $C_{l}^{\phi\phi}$ and the matter power spectrum $P_{\text{m}}\left(k\right)$
are plotted with the chosen values of $\alpha_{\text{H}}=0$ and $\beta_{3}=1\times10^{-4}$.
The colored lines correspond to different values of the effective
Planck mass, while black dashed line corresponds to the $\Lambda$CDM case. 
\label{Fig_Power_Spetra}}
\end{figure}

Thus far we have primarily focused on gravity, and have not included any matter field other than the DE one excited by the modifications. The stability conditions that we have derived in this Section are therefore derived in vacuum. We leave the derivation of the extended stability criteria when matter is included for future work, and here simply mention that in general there will be a mixing of degrees of freedom which would modify the stability conditions. Nevertheless, before concluding we would like to run a preliminary analysis of the cosmological phenomenology of our model by looking at the CMB temperature and weak lensing angular power spectra as well as at the matter power spectrum. In order to do so, we implicitly add matter in the model, minimally coupling it to gravity. At the background level, the matter sectors simply contribute to the RHS of the Friedmann equations as external sources, but they do not impact our analysis of the DE dynamics. At linear level, the matter sector will contribute to the kinetic matrix of the quadratic action for scalar perturbations, requiring a de-mixing which affects the resulting speed of sound of DE \citep{Kase:2014aa,Felice:2016aa,Aoki:2025aa}. As we just mentioned, this should be taken into account when checking for the stable parameter space of the model, and we leave a complete derivation for future work. Let us however note that in this scenario, the scalar mode of Freezing Gravity will no longer be frozen in the IR limit.

Finally, we use the Einstein-Boltzmann solver \texttt{EFTCAMB} to explore the implications of Freezing Gravity for some cosmological observables. We show the predictions for the CMB temperature and weak lensing angular power spectra, as well as the matter power spectrum in Fig. \ref{Fig_Power_Spetra}. We will leave a thorough comparison with the currently available cosmological data for future work. However, we can see that, in general, the predicted cosmology is consistent with the $\Lambda$CDM model, while displaying some interesting, mild deviations.

\section{Conclusions \label{sec_Conclusions}}

In this work, we have proposed a general model, dubbed Freezing Gravity (\ref{Sec-III-B_S_PCDE}), for dark energy (DE) that is capable of realizing arbitrary evolutions of the cosmological background while avoiding the ghost and gradient instabilities in the linear perturbations. Freezing Gravity is constructed within the framework of spatially covariant gravity with the velocity of lapse function (SCG-$L$) theories \citep{Gao2019c}. The equation of state (EoS) of DE (\ref{Sec-III_w_de}) is directly determined by two free functions of time, namely $\left\{\mathcal{F},\mathcal{G}\right\}$ from the Lagrangian (\ref{Sec-III-B_S_PCDE}) and thus independent of the dynamics of the fields, which allows one to reproduce any desired behavior of EoS, including the phantom crossing ones, without needing to solve the background equations of motion (EoM) (\ref{Sec-III-B_beta=00003D0_EoM_BG}). Perturbations  are independently controlled by another set of coefficients, i.e. $\left\{\sigma_{1},\rho_{1},\rho_{2}\right\}$ which are the generic functions of both time and lapse, making the stability of perturbations independent of the solutions for the background.

At linear level in the perturbations, 
after integrating out the auxiliary field, we obtain the effective action for the curvature perturbation, $\zeta$, and show that in the long-wavelength limit the kinetic term approaches zero, seemingly indicating that the model becomes strongly coupled in the IR regime. However, we speculate that the strong coupling issue is circumvented because the kinetic terms are vanishing not only at the linear order but also at any higher order in the IR limit. We demonstrate that this is true up to cubic order by showing that the dangerous terms (\ref{Sec-IV-B_L_3}) appearing in the cubic action do vanish in the IR limit. Given this characteristic behavior on large scales, we dub the model \emph{Freezing Gravity}. In the short-wavelength limit, the kinetic term changes sign when the free parameter $\alpha_{\text{L}}$ varies in $\left(-1,+\infty\right)$, signaling the energy cut-off scale of the model. In fact, in the case of $-1<\alpha_{\text{L}}<0$, there is a pole in the sound speed  which determines the energy cut-off scale of the model and yields an upper bound to $\beta_{3}$ (\ref{Sec-IV-C_beta_3_ub}). The ghost instability is avoided below the energy cut-off scale and we identify the criteria for avoiding the gradient instability. The case of $\alpha_{\text{L}}\geq0$ is free from ghost instability while the gradient stability imposes another set of conditions. 

We use the Einstein-Boltzmann solver \texttt{EFTCAMB} to explore the cosmological phenomenology of stable Freezing Gravity models and compare it to that of $\Lambda$CDM; we show that in general they can produce consistent results with some interesting deviations also within current observational bounds. 

We conclude with several remarks. First, the recent potential indication of phantom crossing behavior of the DE sector is based entirely on cosmological background observations, i.e. DESI BAO and supernova luminosity distance measurements, but has often been associated with theoretical stability of perturbations. Freezing Gravity provides an explicit example where perturbations are detached from the background in the sense that an arbitrary background evolution, crossing the phantom divide or not, can be realized by choosing appropriate $\{\mathcal{F},\mathcal{G}\}$ in eq.\eqref{Sec-III-B_S_PCDE}. While the stability of perturbations is controlled by a completely different set of parameters $\{\sigma_1,\rho_1,\rho_2\}$, making it possible to stabilize on arbitrary background evolution. This reveals a fundamental difficulty in constraining theory with background-only observations and highlights the importance of observations probing the perturbations.  

Our main focus was the dynamics of DE, thus when we studied the stability properties of our model, we did so in the absence of any other matter field. It is well known that the inclusion of matter will in general modify the stability conditions \citep{Kase:2014aa,Felice:2016aa} (also see \citep{Aoki:2025aa}), because of a mixing of degrees of freedom. We plan to derive the complete stability conditions in the presence of matter in a future work.

Let us also comment on the null energy condition (NEC) violation and non-singular cosmology solutions in Freezing Gravity. It is well known that stable nonsingular cosmologies are prohibited within Horndeski theory if the geodesic completeness for gravitons is required \citep{Libanov2016,Kobayashi2016}, while it is achievable in theories beyond Horndeski \cite{Cai:2016ab,Creminelli:2016zwa,Kolevatov:2017aa} (see also \cite{Ye:2025ab}).
Apparently, stable NEC violation can be achieved in FG, even if it is reduced to the subclass of Horndeski theory. This is because in this case the scalar DoF of Freezing Gravity is completely suppressed and associated to the cuscuton theory where the No-Go theorem does not apply. 
 
Finally, we arrived at the action of Freezing Gravity obtained by imposing the conditions (\ref{Sec-III-B_D_eqs}) and (\ref{Sec-III-B_rho_0}) on the free functions of the action, which enable an algebraic phantom crossing condition and make the EoS of DE independent of background solutions. These requirements tightly restrict the theoretical space. An alternative worth exploring in future work, is to impose conditions on the background EoMs instead, such that they can be generally solved for $H$ and $\bar{N}$.\footnote{Another example whose general solution to its Friedmann equations is obtainable was recently proposed in \cite{Hell:2025lgn}} Then one would be able to obtain the EoS of DE explicitly by substituting the general solutions into the expression for $w_{\textrm de}$, with  less constraining limitations of the theoretical space. We leave this for future work.
 
\appendix

\section{Map between U-DHOST and SCG-$L$ theories} \label{App_UDHOST_SCGL}

The relations between the two sets of coefficients in actions
(\ref{Sec-II_S_U-DHOST}) and (\ref{Sec-II_S}) are the following
\begin{equation}
\alpha=\frac{1}{N^{2}}\left(B_{1}-\frac{1}{N^{2}}B_{2}\right)+\frac{1}{N}\left(B_{1}+2\dot{G}\right)\gamma,\quad\gamma=\frac{3\left(C_{3}-2N^{2}C_{2}-2N^{5}G_{N}\right)}{2N^{3}\left(C_{1}+3C_{2}-2N^{2}G\right)},\label{SCGL_DHOST_Coeffieciet_1}
\end{equation}
\begin{equation}
\beta=-\frac{1}{N}\left(B_{1}+2\dot{G}\right),\quad\sigma_{1}=G+\frac{1}{N^{2}}C_{1},\quad\sigma_{2}=\frac{1}{N^{2}}\left(\frac{1}{3}C_{1}+C_{2}\right)-\frac{2}{3}G,\label{SCGL_DHOST_Coeffieciet_2}
\end{equation}
and
\begin{equation}
\rho_{0}=P\left(t,\frac{1}{2N^{2}}\right),\quad\rho_{1}=G\left(t,\frac{1}{2N^{2}}\right),\quad\rho_{2}=\frac{1}{N^{2}}\left(\frac{1}{N^{2}}C_{4}-C_{1}\right)+2NG_{N}.\label{SCGL_DHOST_Coeffieciet_3}
\end{equation}
where the coefficients in the RHS should be understood as functions of $\left(t,\frac{1}{2N^{2}}\right)$.

The inverse maps can be found as follows
\begin{equation}
P=\rho_{0}\left(\phi,\frac{1}{\sqrt{2X}}\right),\quad G=\rho_{1}\left(\phi,\frac{1}{\sqrt{2X}}\right),
\end{equation}
\begin{equation}
B_{1}=-\left(\frac{1}{\sqrt{2X}}\beta+2\rho_{1\phi}\right),\quad B_{2}=-\frac{1}{2X}\left[\frac{1}{2X}\left(\alpha+\beta\gamma\right)+\frac{1}{\sqrt{2X}}\beta+2\rho_{1\phi}\right],
\end{equation}
\begin{equation}
C_{1}=\frac{1}{2X}\left(\sigma_{1}-\rho_{1}\right),\quad C_{3}=\frac{1}{2X^{2}}\left[\left(1+\frac{\gamma}{\sqrt{2X}}\right)\sigma_{2}-\frac{1}{3}\sigma_{1}+\rho_{1}+\frac{1}{\sqrt{2X}}\rho_{1X}\right],
\end{equation}
and
\begin{equation}
C_{2}=\frac{1}{2X}\left(\sigma_{2}-\frac{1}{3}\sigma_{1}+\rho_{1}\right),\quad C_{4}=\frac{1}{2X^{2}}\left(\sigma_{1}+\frac{1}{2}\rho_{2}-\rho_{1}-\frac{1}{\sqrt{2X}}\rho_{1X}\right),
\end{equation}
where the coefficients in the RHS should be understood as functions of $\left(\phi,\frac{1}{\sqrt{2X}}\right)$.

\section{DE energy density and pressure} \label{App_energy_pressure_DE}

The coefficients in (\ref{Sec-III-A_rho_de}) and (\ref{Sec-III-A_p_de}) are given as follows
\begin{eqnarray}
\mathcal{W}_{\dot{H}} & \equiv & 6\bar{N}\gamma\sigma_{2},\quad\mathcal{W}_{H^{2}}\equiv3\left[3\left(2\bar{N}\gamma+1\right)\sigma_{2}-3\bar{N}\partial_{N}\sigma_{2}+2\right],\\
\mathcal{W}_{\dot{F}} & \equiv & 2\bar{N}\gamma^{2}\sigma_{2},\quad\mathcal{W}_{F^{2}}\equiv\bar{N}\gamma^{2}\sigma_{2}\left(\partial_{N}\ln\sigma_{2}+2\partial_{N}\ln\gamma+\frac{1}{\bar{N}}\right),\\
\mathcal{W}_{H} & \equiv & 3\left[2\partial_{t}\left(\gamma\sigma_{2}\right)+\bar{N}\left(\alpha+\beta\gamma-\partial_{N}\beta\right)\right],\quad\mathcal{W}_{HF}\equiv6\gamma\sigma_{2}\left(\bar{N}\gamma+1\right),\\
\mathcal{W}_{F} & \equiv & 2\gamma\left(\gamma\partial_{t}\sigma_{2}+2\sigma_{2}\partial_{t}\gamma\right),\quad\mathcal{W}_{N}\equiv\partial_{t}\left(\alpha+\beta\gamma\right)-\partial_{N}\left(\bar{N}\rho_{0}\right),
\end{eqnarray}
and
\begin{eqnarray}
\mathcal{M}_{\dot{H}} & \equiv & 6\bar{N}\left(3\sigma_{2}+2\right),\quad\mathcal{M}_{\dot{F}}\equiv\bar{N}\gamma\sigma_{2},\\
\mathcal{M}_{H^{2}} & \equiv & 9\bar{N}\left(3\sigma_{2}+2\right),\quad\mathcal{M}_{HF}\equiv18\bar{N}\partial_{N}\sigma_{2},\\
\mathcal{M}_{F^{2}} & \equiv & 6\bar{N}\gamma\sigma_{2}\left(\partial_{N}\ln\sigma_{2}+\partial_{N}\ln\gamma-\frac{1}{2}\gamma\right),\\
\mathcal{M}_{H} & \equiv & 18\partial_{t}\sigma_{2},\quad\mathcal{M}_{N}\equiv3\left(\partial_{t}\beta-\bar{N}\rho_{0}\right),\\
\mathcal{M}_{F} & \equiv & 3\left[2\partial_{t}\left(\gamma\sigma_{2}\right)-\bar{N}\left(\alpha+\beta\gamma-\partial_{N}\beta\right)\right].
\end{eqnarray}

\section{EFT action in the $M_{i}$ functions}\label{App_M_functions}

According to the map given in \citep{Frusciante:2016aa}, the quadratic EFT action of Freezing Gravity (\ref{Sec-II_S}) can be written in terms of the $M_{i}$ functions as follows
\begin{eqnarray}
S_{2}^{\text{EFT-}M} & = & \int\text{d}^{4}x\sqrt{-g}\Big\{\frac{M_{\text{Pl}}^{2}}{2}\left[1+\Omega\left(t\right)\right]\mathscr{R}+\Lambda\left(t\right)-c\left(t\right)\delta g^{00}\nonumber \\
 &  & -\frac{\bar{M}_{1}^{3}\left(t\right)}{2}\delta g^{00}\delta K-\frac{\bar{M}_{2}^{2}\left(t\right)}{2}\left(\delta K\right)^{2}-\frac{\bar{M}_{3}^{2}\left(t\right)}{2}\delta K^{\mu}{}_{\nu}\delta K^{\nu}{}_{\mu}\nonumber \\
 &  & +\frac{M_{2}^{4}\left(t\right)}{2}\left(\delta g^{00}\right)^{2}+\frac{\hat{M}^{2}\left(t\right)}{2}\delta g^{00}\delta R+m_{2}^{2}\left(t\right)h^{\mu\nu}\partial_{\mu}\delta g^{00}\partial_{\nu}\delta g^{00}\Big\},
\end{eqnarray}
where the mass parameters are given by
\begin{eqnarray}
\Omega\left(t\right) & = & \frac{2\rho_{1}}{M_{\text{Pl}}^{2}}-1,\quad c\left(t\right)=H\dot{\rho}_{1}-\ddot{\rho}_{1}-2\rho_{1}\dot{H},\\
\Lambda\left(t\right) & = & -2\left(3H^{2}\rho_{1}+\ddot{\rho}_{1}+2H\dot{\rho}_{1}+2\rho_{1}\dot{H}\right),\quad M_{2}^{4}\left(t\right)=-\frac{1}{2}c,\\
\hat{M}^{2}\left(t\right) & = & \partial_{N}\rho_{1},\quad m_{2}^{2}\left(t\right)=\frac{1}{4}\rho_{2},\quad\bar{M}_{1}^{3}\left(t\right)=-2\dot{\rho}_{1},\\
\bar{M}_{2}^{2}\left(t\right) & = & \frac{2}{3}\left(\sigma_{1}+2\right)-2\rho_{1},\quad\bar{M}_{3}^{2}\left(t\right)=2\left(\rho_{1}-\sigma_{1}\right).
\end{eqnarray}

\section{Coefficients of the effective quadratic action\label{App_gradient_coeff}}

The $\mathcal{B}_{0}$, $\mathcal{B}_{2}$ and $\mathcal{B}_{4}$ in eq. (\ref{Sec-IV-B_A=000026B}) are given as follows 
\begin{eqnarray}
\mathcal{B}_{0} & \equiv & \left(\mathcal{B}_{03}\dot{H}+\mathcal{B}_{02}H^{2}+\mathcal{B}_{01}H\right)H^{2},\label{App_B_B_0}\\
\mathcal{B}_{03} & \equiv & -72\left(\alpha_{\text{L}}+1\right)^{2}\left(\alpha_{\text{H}}+1\right),\\
\mathcal{B}_{02} & \equiv & 72\left(\alpha_{\text{L}}+1\right)^{2}\left(\alpha_{\text{H}}-\alpha_{\text{T}}\right),\\
\mathcal{B}_{01} & \equiv & 72\left(\alpha_{\text{L}}+1\right)^{2}\left(\alpha_{\text{H}}+1\right)\frac{\text{d}}{\text{d}t}\ln\left(\frac{\alpha_{\text{H}}+1}{\alpha_{\text{L}}+1}\right),
\end{eqnarray}
and
\begin{eqnarray}
\mathcal{B}_{2} & \equiv & \mathcal{B}_{23}\dot{H}+\mathcal{B}_{22}H^{2}+\mathcal{B}_{21}H,\label{App_B_B_2}\\
\mathcal{B}_{23} & \equiv & 12\beta_{3}\alpha_{\text{L}}\left(\alpha_{\text{L}}+1\right)\left(\alpha_{\text{H}}+1\right),\label{App_B23}\\
\mathcal{B}_{22} & \equiv & 24\alpha_{\text{L}}\left(\alpha_{\text{L}}+1\right)\left[\left(\alpha_{\text{H}}+1\right)^{2}+\beta_{3}\left(\frac{3}{2}\alpha_{\text{H}}-\alpha_{\text{T}}\right)\right],\\
\mathcal{B}_{21} & \equiv & 12\beta_{3}\alpha_{\text{L}}\left(\alpha_{\text{L}}+1\right)\left(\alpha_{\text{H}}+1\right)\frac{\text{d}}{\text{d}t}\ln\left(\frac{\alpha_{\text{H}}+1}{\alpha_{\text{L}}\beta_{3}}\right),
\end{eqnarray}
with
\begin{equation}
\mathcal{B}_{4}\equiv2\beta_{3}\alpha_{\text{L}}^{2}\left[2\left(\alpha_{\text{H}}+1\right)^{2}-\beta_{3}\left(\alpha_{\text{T}}+1\right)\right].\label{App_B_B_4}
\end{equation}

\acknowledgments
Z. Yao would like to thank Mohammad Ali Gorji for useful discussions. We acknowledge support from the European Research Council under the H2020 ERC Consolidator Grant “Gravitational Physics from the Universe Large scales Evolution” (Grant No. 101126217 — GraviPULSE) and from the NWO and the Dutch Ministry of Education, Culture and Science (OCW) (through NWO VIDI Grant No. 2019/ENW/00678104 and ENW-XL Grant OCENW.XL21.XL21.025 DUSC).

\end{document}